\documentstyle[seceq,epsbox,twocolumn]{jpsj}

\def\runtitle{Application of the CVM to Spin Ice Systems}
\def\runauthor{Shun-ichi \textsc{Yoshida} ,\, Koji \textsc{Nemoto} ,\, Koh \textsc{Wada}}

\title{Application of the Cluster Variation Method to
Spin Ice Systems on the Pyrochlore Lattice}
\author{Shun-ichi \textsc{Yoshida}\thanks{E-mail: shun1@statphys.sci.hokudai.ac.jp} ,\, Koji \textsc{Nemoto} ,\, Koh \textsc{Wada}}
\inst{
Division of Physics, Graduate School of Science,
 Hokkaido University, Sapporo 060-0810
}

\recdate{\today}

\abst{
The cactus approximation in the cluster variation method
 is applied to the spin ice system
  with nearest neighbor ferromagnetic coupling.
The temperature dependences of the entropy and the specific heat
 show qualitatively good agreement with those observed
  by Monte Carlo simulations and experiments, and the Pauling value
   is reproduced for the residual entropy.
The analytic expression of
 the $\mib{q}$-dependent magnetic susceptibility is obtained, from which 
  the absence of magnetic phase transition is confirmed.
The neutron scattering pattern is also evaluated and found to be
 consistent with that obtained from Monte Carlo simulations.
}

\kword{
Spin Ice, Geometrical Frustration System, Cluster Variation Method, Cactus Approximation.
}

\begin{document}
\sloppy
\maketitle

\section{Introduction}

\begin{figure}[tbp]
{\scriptsize (a)}\begin{center}
\epsfile{file=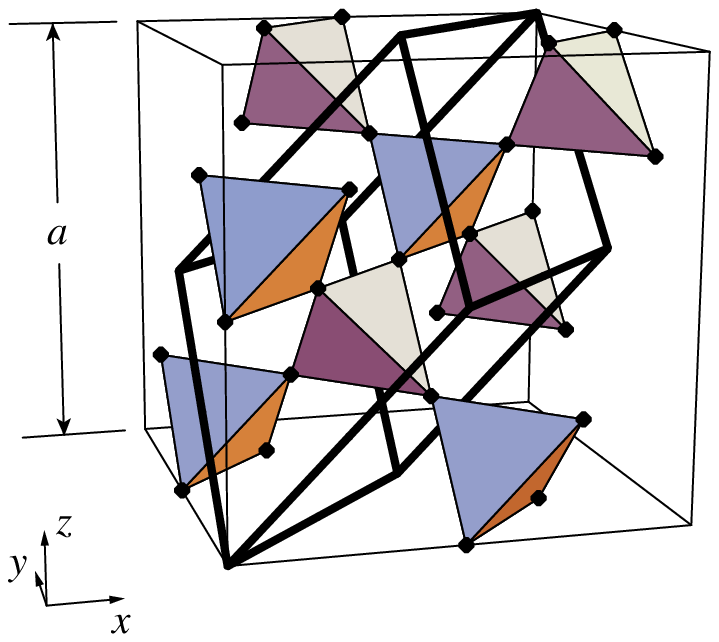,width=7cm}
\end{center}
\medskip
{\scriptsize (b)}\begin{center}
\epsfile{file=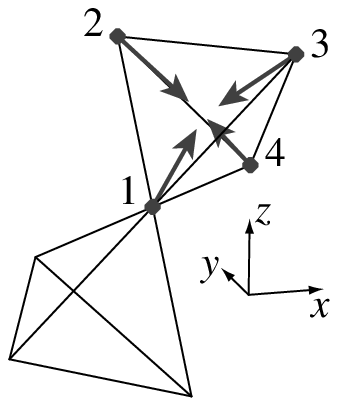,width=7cm}
\end{center}
\medskip
{\scriptsize (c)}\begin{center}
\epsfile{file=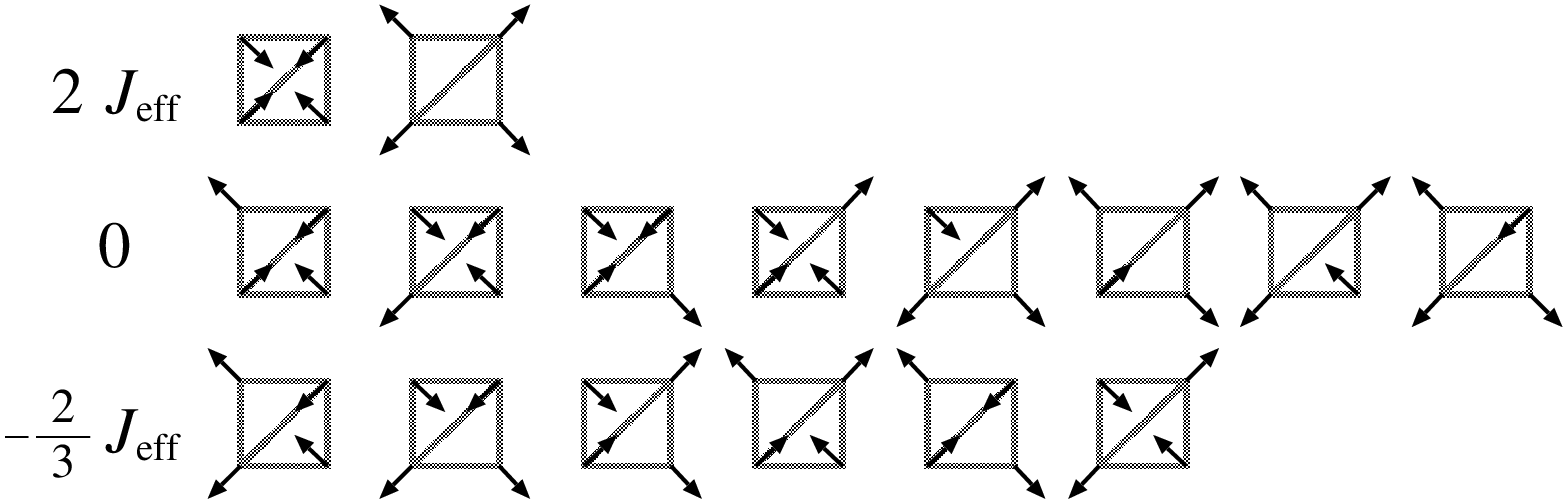,width=8.5cm}
\end{center}
\caption{
(a) A schematic plot of the pyrochlore lattice. 
The magnetic ions are located at each vertex of the tetrahedra. 
The outlines of the rhombohedral chemical unit-cell and the cubic
 unit-cell are also shown. 
There are four independent spins and two tetrahedra in a rhombohedral
 chemical unit-cell.
(b) Local Ising axes $\mib{n}_\nu \,(\nu=1,\,2,\,3,\,4)$
(c) Energy levels of spin configurations in one tetrahedron. 
}
\label{fig:pyrochlore}
\end{figure}

In recent years the low temperature physics of `spin ice' systems
 such as Ho$_2$Ti$_2$O$_3$ and Dy$_2$Ti$_2$O$_3$ draws
  a great deal of attentions of many researchers.
Among them the most remarkable property is that such materials have a
 residual entropy whose value is almost the same as that obtained by
  Pauling\cite{Pauling} for proton configurations in the ordinary
   hexagonal ice I$_{\rm h}$. 
This indicates the existence of the `ice rule' 
 in these magnetic materials\cite{Harris1997} due to spin frustrations.
The difference of `spin ice' from the real ice is that we can resolve
 the degeneracy by applying an infinitesimal magnetic field.
One of our interests then is how the spins behave in response to
 magnetic fields.
The spin ice system has a spin lattice forming the pyrochlore structure
 as depicted in Fig.~\ref{fig:pyrochlore}(a).
In the pyrochlore lattice the spin sites construct a three dimensional
 network of corner-sharing tetrahedra with cubic symmetry.
Each site belongs to two tetrahedra and the spin is forced to point
 toward either of these centers by strong crystal field anisotropy. 
Due to this Ising-like property the spin system becomes fully-frustrated
 on the pyrochrore lattice if the neighboring spins are likely to align
  ferromagnetically because the direction-cosine between their easy-axes
   is negative.
In this case the lowest energy of four-spin state on a tetrahedron is
 given by the configurations
  where two spins point ``in'' and the other two spins ``out'',
   and it is six-fold degenerated (see Fig.~\ref{fig:pyrochlore}(c)).
This rule applied in the ground state is
 the same as that for the proton configuration in the ordinary hexagonal
  ice I$_{\rm h}$, and is therefore called the ice rule.
The ice rule makes the number of low-lying states of the whole system
 enormous as long as the long range spin-spin interactions are very
  weak.
In particular the ground state is macroscopically degenerated
 when only the nearest neighbor interaction is concerned.
Pauling is the first one who estimated the value of the residual entropy
 to obtain
\begin{equation}
 \frac{S_0}{Nk_{\rm B}}
 =\frac{\,1\,}{\,2\,}\ln\frac{\,3\,}{\,2\,}
 \simeq 0.203,
\label{PaulingValue}
\end{equation}
  where $N$ is the number of protons in the ice I$_{\rm h}$.
Although his original interest was the entropy due to
 the proton configurations in the ordinary hexagonal ice I$_{\rm h}$,
  the value obtained is the same for the pyrochlore lattice since the
   network topology of tetrahedra is ignored in his estimation.
It turned out that
 the difference between the structure of tetrahedra
   in the pyrochlore and I$_{\rm h}$ lattices
    is not so relevant because the correction to the Pauling value
      is very small.~\cite{Nagle1965} 
This is the reason why the Pauling value is a good approximate one for
 both systems.
Indeed the specific heat for spin ice systems was measured
 by experiments and by Monte Carlo simulations,~\cite{Hertog2000} from which 
  the residual entropy is evaluated and agrees with (\ref{PaulingValue}).
A mean field approximation (MFA) may be the first step
 to examine thermodynamic properties and has been so far applied
  to the spin ice system to analyze neutron scattering
   experiments.~\cite{Reimers1991}\cite{Kadowaki2001} 
The simplest MFA is the one using the 4-sublattice magnetizations
 and neglecting spin correlations.
Such a treatment, however, results in a phase transition at a finite
 temperature, below which a long range order appears and brings about
  the vanishing of the entropy as $T\rightarrow0$.~\cite{Reimers1991}
This discrepancy from those of experiments and Monte Carlo simulation is
 due to the fact that the MFA cannot deal with the spin fluctuations and
  ``two-in and two-out'' spin behaviors in tetrahedra, i.e., the ice rule.

On the other hand, Slater studied a hydrogen bonded 
 crystal KH$_2$PO$_4 \rm{(KDP)}$~\cite{Slater1941} with the ice rule.
In this crystal the ice rule is described by that 
(i) one proton exists on each O-O bond between two nearest neighboring
 PO$_4$  tetrahedra and 
(ii) each tetrahedron has exactly two protons ajacent to it.
He succeded in explaining the phase transition of KDP by coping with the
 ice rule.
His approximate treatment is tantamount to an extended Bethe
 approximation. 
It is equivalent to the tetrahedral cactus approximation in the cluster
 variation method (CVM), which was developed to improve the MFA
  systematically by taking into account higher order
   correlations.~\cite{Kikuchi1951}
The tetrahedral cactus approximation in the CVM was successfully applied
 to analyze the wave-number dependent ferro-electric susceptibility
  above and below the transition temperature for the KDP
   crystal.~\cite{Wada1998}
 
In the present paper, we apply the tetrahedral cactus approximation in
 the CVM to the spin ice system with nearest-neighbor ferromagnetic
  coupling.
Within the present approximation we show that no magnetic phase
 transition takes place by investigating analytically obtained
  eigenvalues appearing in the wave-number dependent susceptibility. 
As a result the Pauling value for the residual entropy is reproduced
 in the zero temperature limit. 
Furthermore we evaluate the neutron scattering intensity,
 which is qualitatively in good agreement with that obtained by
  Monte Carlo simulations and experiments.

The present manuscript is organized as follows: 
In \S\ref{sec:formulation} we present the CVM formulation for the spin
 ice system with nearest neighbor ferromagnetic coupling to derive a set
  of linear response equations to an inhomogeneous magnetic field. 
In \S\ref{sec:results} described  are the main results 
 including the temperature dependency of the entropy and of the specific
  heat, the wave-number dependent susceptibility and the neutron
   scattering pattern.
Section \ref{sec:discussion} is devoted to a summary of this manuscript
 with some discussions on the relation between the present
  approximation and the usual mean field approximation.

\def\vS{\mib{S}}
\def\vH{\mib{H}}
\def\vr{\mib{r}}
\def\vR{\mib{R}}
\def\vn{\mib{n}}
\def\braket#1{{\langle #1 \rangle}}
\def\Jeff{J_{\mbox{{\scriptsize eff}}}}
%%%%%%% section 2 %%%%%%%%%
\section{Formulation of Nearest Neighbor Spin Ice
 System}\label{sec:formulation}
\subsection{Indexing Site Locations}\label{subsec:indexing}
The pyrochlore lattice is a non-Bravias lattice. 
In a rhombohedral chemical unit cell located at a translational lattice
 site $\vR$ there exist four independent spins situated on a tetrahedron 
  whose position vectors are denoted by $\vr_\nu$ relative to $\vR$. 
Thus there are four sublattices for spin configurations in the
pyrochlore lattice.
For $N/4$ rhombohedral chemical unit cells the translational lattice
 vector $\vR$ is chosen in the center of each unit cell.  
The $\nu$-th position vector in the unit cell is taken, respectively, as
\begin{eqnarray}
&&\mib{r}_1
 = 
\frac{\,a\,}{\,4\,}\left(\!\!
\begin{array}{c}
 0\\
 0\\
 0
\end{array}\!\!
\right)\,,\quad
\mib{r}_2
 = 
\frac{\,a\,}{\,4\,}\left(\!\!
\begin{array}{c}
 0\\
 1\\
 1
\end{array}\!\!
\right)\,,\quad
\nonumber\\[-2mm]
\\[-2mm]
&&\mib{r}_3
 = 
\frac{\,a\,}{\,4\,}\left(\!\!
\begin{array}{c}
 1\\
 0\\
 1
\end{array}\!\!
\right)\,,\quad
\mib{r}_4
 = 
\frac{\,a\,}{\,4\,}\left(\!\!
\begin{array}{c}
 1\\
 1\\
 0
\end{array}\!\!
\right)\,,\quad
\nonumber
\end{eqnarray}
where $a$ is the linear size of the cubic primitive cell (see
 Fig.\ref{fig:pyrochlore}(a)). 
For later notational convenience we sometimes use the sequential
 numbering $i$-th site $(i=1,\cdots,N)$ for a spin site $\vR+\vr_\nu$. 
%###############
\subsection{Spin Hamiltonian}
The Hamiltonian of the spin ice system with $N$ spins under an
 inhomogeneous external field $\vH_i$ ($i=1,\,\cdots,\,N$) is given as
\begin{eqnarray}
\lefteqn{{\cal H}(\vS_1,\,\cdots,\,\vS_N)}\nonumber\\ 
 &=& -D_{\rm a} \sum_i (\vn_{\nu(i)}\cdot\vS_i)^2
  -J\sum_\braket{ij}\vS_i \cdot \vS_j
  -\sum_i \vH_i \cdot \vS_i \nonumber\\[-2mm]
  && + D r_{\rm nn}^3 \sum_\braket{ij}
        \left[
                \frac{\vS_i\cdot\vS_j}{|\vr_{ij}|^3}
                -\frac{3(\vS_i\cdot\vr_{ij})(\vS_j\cdot\vr_{ij})}{|\vr_{ij}|^5}
        \right]\,,
\label{eqn:Hamiltonian}
\end{eqnarray}
where $\vS_i$ is a spin vector at the $i$-th site $(|\vS_i|=1)$, 
$\vr_{ij}$ the displacement vector from the $j$-th to the $i$-th
 site, and $r_{\rm nn}$ is the distance
  between nearest neighboring two spins.
The second term represents
 the nearest-neighbor ferromagnetic exchange interactions $(J>0)$.
The third and the last terms are the Zeeman energy and  
 the dipolar interactions, respectively. 
The first term,
 which represents the single-ion anisotropy($D_{\rm a}>0$),
  is the most crucial one in the spin ice system because this term makes
   the system obey the ice rule.
There are four distinct local easy-axis
   directions denoted by $\vn_{\nu(i)}$,
    where $\nu$ denotes the sublattice where the $i$-th spin belongs.
As shown in Fig.~\ref{fig:pyrochlore}(b)
 we take the unit vector $\vn_\nu$ directing the local Ising axis
  of the $\nu$-th sublattice as
\begin{eqnarray}
&&\mib{n}_1
 = 
\frac{\,1\,}{\,\sqrt{3}\,}\left(\!\!
\begin{array}{r}
 1\\
 1\\
 1
\end{array}\!\!
\right)\,,\quad
\mib{n}_2
 = 
\frac{\,1\,}{\,\sqrt{3}\,}\left(\!\!
\begin{array}{r}
 1\\
 -1\\
 -1
\end{array}\!\!
\right)\,,\quad
\nonumber\\[-2mm]
\\[-2mm]
&&\mib{n}_3
 = 
\frac{\,1\,}{\,\sqrt{3}\,}\left(\!\!
\begin{array}{r}
 -1\\
 1\\
 -1
\end{array}\!\!
\right)\,,\quad
\mib{n}_4
 = 
\frac{\,1\,}{\,\sqrt{3}\,}\left(\!\!
\begin{array}{r}
 -1\\
 -1\\
 1
\end{array}\!\!
\right)\,.\quad
\nonumber
\end{eqnarray}

Now we introduce two assumptions to make the model a little simpler.
First we consider the case where the dipole interactions are so weak
 $(D\ll J)$ that they are negligible except the nearest-neighbor one,
  although one has to cope with the effect of $D$ when expecting
   that the system falls into the unique ground state in very low
    temperatures.
Second the single-ion anisotropy is so strong $(D_{\rm a}\gg J)$ that
 the spins are forced to align along their uniaxial directions.
Then each of the spins $\vS_i$ can be expressed
 with an Ising variable $\sigma_i$ as
\begin{equation}
 \mib{S}_i = \sigma_i \mib{n}_{\nu(i)}\,,\quad \sigma_i=\pm1\,,
\end{equation}
and the Hamiltonian (\ref{eqn:Hamiltonian})
 can be expressed in terms of $\{\sigma_i\}$ as
\begin{equation}
 {\cal H}(\sigma_1,\,\cdots,\,\sigma_N)
 \simeq \frac{\Jeff}{3}\sum_\braket{ij}\sigma_i\sigma_j
  -\sum_i H_i\sigma_i\,,
\end{equation}
where $\Jeff=J - 5 D(>0)$
 and $H_i=\vH_i\cdot\vn_{\nu(i)}$. 
Note that the interaction term is antiferromagnetic,
which means that the system is fully-frustrated.

For the convenience of the following approximation, it is useful to
 divide the above Hamiltonian into those for the tetrahedra:
\begin{equation}
 {\cal H}(\sigma_1,\,\cdots,\,\sigma_N)
  = \sum_\braket{i j k l}
        {\cal H}^{(ijkl)}_4(\sigma_i,\,\sigma_j,\,\sigma_k,\,\sigma_l),
\label{eqn:sumH4}
\end{equation}
where  the superscript $(ijkl)$ denotes a set of four spin sites
 on a tetrahedron and the index $\braket{ijkl}$ runs over all tetrahedra. 
The Hamiltonian in a tetrahedron ${\cal H}^{(ijkl)}_4$ consists of
 the interaction term ${\cal H}_4^0$ and the Zeeman term
 ${\it \Delta}{\cal H}^{(ijkl)}_4$ : 
\begin{eqnarray}
\lefteqn{\hspace{-0.7cm}{\cal H}^{(ijkl)}_4 (\sigma_i,\,\sigma_j,\,\sigma_k,\,\sigma_l)}\hspace{-0.5cm}&& \label{eqn:H4}\\
 && = {\cal H}_4^0 (\sigma_i,\,\sigma_j,\,\sigma_k,\,\sigma_l)
   + {\it\Delta}{\cal H}^{(ijkl)}_4 (\sigma_i,\,\sigma_j,\,\sigma_k,\,\sigma_l),\nonumber
\end{eqnarray}
where
\begin{eqnarray}
&&{\cal H}_4^0 (\sigma_i,\,\sigma_j,\,\sigma_k,\,\sigma_l)\quad \nonumber\\
&&\quad = \frac{J_{\mbox{{\scriptsize eff}}}}{\,3\,} (\sigma_i \sigma_j + \sigma_j \sigma_k + \sigma_k \sigma_l + \sigma_l \sigma_i + \sigma_i \sigma_k + \sigma_j \sigma_l)\,,\nonumber\\[-1mm]
\label{eqn:H40}\\[2mm]
&&{\it \Delta}{\cal H}^{(ijkl)}_4(\sigma_i,\,\sigma_j,\,\sigma_k,\,\sigma_l)\nonumber\\
&&\qquad = - \frac{\,1\,}{\,2\,} (H_i \sigma_i + H_j \sigma_j + H_k \sigma_k + H_l \sigma_l)\,.\label{eqn:DeltaH4}
\end{eqnarray}

\def\Ptot{P_{\rm tot}}
\subsection{Cactus Approximation in the CVM}
The starting point of the cluster variation method (CVM) is the
 following well-known variation principle.
Let 
$\Ptot(\sigma_1,\,\cdots,\,\sigma_N)$
 be the probability of a whole spin configuration.
Then the  variational free energy of the system $F$ is given by
\begin{eqnarray}
 \beta F
 &=& \mbox{\large\rm Tr}\bigl[\beta {\cal H}(\sigma_1,\,\cdots,\,\sigma_N)
         \Ptot(\sigma_1,\,\cdots,\,\sigma_N)\bigr]\label{eqn:free-energy}\\
 && \quad + \mbox{\large\rm Tr}\bigl[
        \Ptot(\sigma_1,\,\cdots,\,\sigma_N) \ln\Ptot(\sigma_1,\,\cdots,\,\sigma_N)\bigr]\,,\nonumber
\end{eqnarray}
where $\beta = 1/k_{\rm B}T$.
The equilibrium probability is obtained by minimizing $F$ with respect to
 $\Ptot(\sigma_1,\,\cdots,\,\sigma_N)$.
Then systematic approximations to $\Ptot$ correspond to
 those to the free energy one by one.
For example the `point' approximation
\begin{equation}
\Ptot(\sigma_1,\,\cdots,\,\sigma_N)\simeq\prod_i P^{(i)}_1(\sigma_i),
\label{eqn:point}
\end{equation}
corresponds to the usual mean field approximation where
 $P^{(i)}_1(\sigma_i)$ denotes the single spin probability.
In the present work we adopt the tetrahedral `cactus' approximation, whose
 approximant is written as
\begin{eqnarray}
\lefteqn{\!\!\Ptot(\sigma_1,\,\cdots,\,\sigma_N)}&&\label{eqn:cactus}\\
&& \,\,\simeq \prod_i P^{(i)}_1(\sigma_i) 
 \prod_{\langle ijkl \rangle} \frac{P^{(ijkl)}_4(\sigma_i,\,\sigma_j,\,\sigma_k,\,\sigma_l)}{P^{(i)}_1(\sigma_i) P^{(j)}_1(\sigma_j) P^{(k)}_1(\sigma_k) P^{(l)}_1(\sigma_l)}\,,\nonumber
\end{eqnarray}
where the probability for the four spins 
$P^{(ijkl)}_4(\sigma_i,\sigma_j,\sigma_k,\sigma_l)$ 
is introduced to take into account two spins `in' and two spins
 `out' correlation on a tetrahedron.
Substituting (\ref{eqn:cactus}) into (\ref{eqn:free-energy}) with
 (\ref{eqn:sumH4}) we obtain the following expression for the
  corresponding variational free energy: 
\begin{eqnarray}
 \beta F
 &\simeq& \sum_{\langle ijkl \rangle} \mbox{{\large\rm Tr}} \bigl[ \beta {\cal H}^{(ijkl)}_4(\sigma_i,\,\sigma_j,\,\sigma_k,\,\sigma_l)\nonumber\\[-5mm]
 && \hspace{2.5cm} \times P^{(ijkl)}_4(\sigma_i,\,\sigma_j,\,\sigma_k,\,\sigma_l)\bigr]\nonumber\\
 && -\sum_{i}\mbox{{\large\rm Tr}} \bigl[P^{(i)}_1(\sigma_i) \ln P^{(i)}_1(\sigma_i)\bigr]\nonumber\\
 && + \sum_{\langle ijkl \rangle} \mbox{{\large\rm Tr}}\bigl[P^{(ijkl)}_4(\sigma_i,\,\sigma_j,\,\sigma_k,\,\sigma_l)\label{eqn:cactus-free-energy}\\[-5mm]
 && \hspace{2.5cm} \times\ln P^{(ijkl)}_4(\sigma_i,\,\sigma_j,\,\sigma_k,\,\sigma_l) \bigr]\,.\nonumber
\end{eqnarray}
Since there are normalization conditions for $P_1$ and $P_4$,
these probability functions are not always independent variational
 parameters to each other.
To eliminate these  restrictions we define order parameters as a set of
 independent variables,
\begin{eqnarray}
 m_a &=& \mbox{{\large\rm Tr}}\bigl[\sigma_a P^{(ijkl)}_4(\sigma_i,\,\sigma_j,\,\sigma_k,\,\sigma_l)\bigr]\,,\\
 m_{ab} &=& \mbox{{\large\rm Tr}}\bigl[\sigma_a \sigma_b P^{(ijkl)}_4(\sigma_i,\,\sigma_j,\,\sigma_k,\,\sigma_l)\bigr]\,,\\
 m_{abc} &=& \mbox{{\large\rm Tr}}\bigl[\sigma_a \sigma_b \sigma_c P^{(ijkl)}_4(\sigma_i,\,\sigma_j,\,\sigma_k,\,\sigma_l)\bigr]\,,\\
 m_{ijkl} &=& \mbox{{\large\rm Tr}}\bigl[\sigma_i \sigma_j \sigma_k \sigma_l P^{(ijkl)}_4(\sigma_i,\,\sigma_j,\,\sigma_k,\,\sigma_l)\bigr]\,.
\end{eqnarray}
where $m_a$ and $m_{abc}$ represent long-range order parameters and
$m_{ab}$ and $m_{ijkl}$ short-range order parameters, respectively, and
$a,\,b,\,c \in \{i,\,j,\,k,\,l\}$. 
In terms of these order parameters  we can express the
 probabilities $P_1$ and $P_4$ as 
\begin{eqnarray}
\lefteqn{P^{(i)}_1(\sigma_i)
 =\frac{\,1\,}{\,2\,}(1+m_i \sigma_i)\,,}&&\\
\lefteqn{P^{(ijkl)}_4(\sigma_i,\,\sigma_j,\,\sigma_k,\,\sigma_l)}&& \quad\nonumber\\
&=&\frac{\,1\,}{\,2^4\,} 
\bigl(1
 + {\textstyle\sum_{a}} m_a\sigma_a
 + {\textstyle\sum_{ab}} m_{ab}\sigma_a\sigma_b \\
&&\qquad
 + {\textstyle\sum_{abc}} m_{abc}\sigma_a \sigma_b \sigma_c
 + m_{ijkl}\sigma_i\sigma_j\sigma_k\sigma_l \bigr)\,.\nonumber 
\end{eqnarray}
Now that  the variational free energy is expressed in terms of the order
 parameters, the thermal equilibrium conditions, in other words,
  the equations of state can be given by its extremum condition:
\begin{equation}
 \frac{\partial F}{\partial m_a}=0\,,\quad
 \frac{\partial F}{\partial m_{ab}}=0\,,\quad
 \frac{\partial F}{\partial m_{abc}}=0\,,\quad
 \frac{\partial F}{\partial m_{ijkl}}=0\,.\quad
\label{eqn:dF/dm=0}
\end{equation}

\section{Results from Cactus Approximation}\label{sec:results}

\subsection{Entropy and Specific Heat}
\begin{figure}[tbp]
\begin{center}
 \epsfile{file=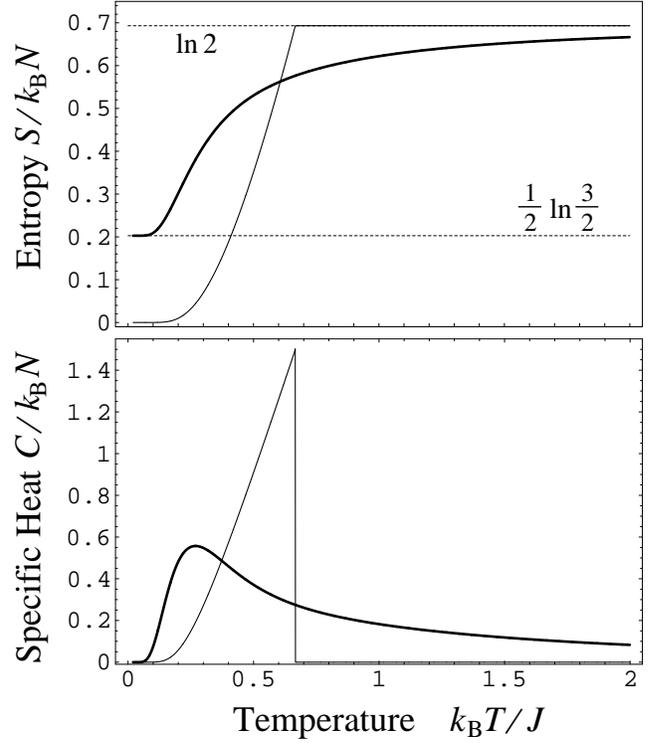,width=8.5cm}
\end{center}
 \caption{Entropy and specific heat versus temperature. 
Entropy at $T=0$ agrees with the Pauling value~(\ref{PaulingValue}). 
Specific heat shows the Schottky anomaly.
Thin curves are results obtained from the mean field approximation. 
}
\label{fig:T-SC}
\end{figure}

In the case of no external field ($H_i=0$) we can derive the analytic
 expression for the probabilities $P_1$ and $P_4$ as
\begin{eqnarray}
&&\hspace{-0.5cm} P^0_1(\sigma) = 1/2 \,,\label{eqn:P01}\\
&&\hspace{-0.5cm} P^0_4(\sigma_i,\,\sigma_j,\,\sigma_k,\,\sigma_l)
  = {\rm e}^{-\beta{\cal H}^0_4(\sigma_i,\,\sigma_j,\,\sigma_k,\,\sigma_l)}/Z\,,\label{eqn:P04}\\
&&\hspace{-0.5cm} Z
 =\mbox{{\large\rm Tr}}\bigl[{\rm e}^{-\beta{\cal H}_4^0(\sigma_i,\,\sigma_j,\,\sigma_k,\,\sigma_l)}\bigr]=6 \eta^{-1} + 8 + 2 \eta^3\,,\label{eqn:Z}
\end{eqnarray}
where $\eta=\exp(-2J_{\mbox{{\tiny eff}}}/3k_{\rm B}T)$ and 
 the superscript ``0'' is used to emphasize thermodynamical equilibrium
  without external field.
In deriving the above solution we assumed that no long range order state 
 happens at finite temperature, which is indeed the case as
  we will see later.
It is straightforward to calculate the entropy $S$ and the specific heat
 $C= T({{\rm d} S}/{{\rm d} T})$  by substituting the above equations in
  the variational free energy (\ref{eqn:cactus-free-energy}) to obtain
\begin{equation}
\frac{S}{Nk_{\rm B}}
 = - \ln 2 + \frac{2J_{\mbox{{\scriptsize eff}}}(\eta^3-\eta^{-1})}{k_{\rm B}T Z} + \frac{\,1\,}{\,2\,}\ln Z\,,
\end{equation}
The temperature dependence of $S$ and $C$ are shown in
 Fig.~\ref{fig:T-SC}. 
We see that the entropy $S$ takes the Pauling value (\ref{PaulingValue})
 in the limit $T\to 0$ and the specfic heat $C$ shows a Schottky anomaly
  as is expected.
These behaviors are in agreement with
 those observed in experiments and in Monte Carlo
  simulations.~\cite{Hertog2000}
For comparison
 the results obtained from the mean field approximation (MFA) are also
  shown by thin curves.
Note that the long range order makes both $S$ and $C$ look very different
 in the low temperature region including $T=0$.

\subsection{Wave-Number dependent Susceptibility}
Here we examine the magnetic property by using the linear response to an 
 inhomogeneous magnetic field $H_i$, in which
  eqs.(\ref{eqn:P01})-(\ref{eqn:Z}) become the bases of linearization.
Some calculations lead us to the linear  equations between 
 magnetization ${\it\Delta}m_i (= m_i)$ and magnetic field.
The detailed derivation is described in appendix\ref{appendix:bH=Ldm} and
 the linear response equation is given by 
\begin{equation}
 \beta H(\mib{R}+\mib{r}_{\nu})
 = \sum_{\nu',\,\mib{R}'}
   {\it\Lambda}_{\nu\nu'} {\it\Delta}m(\mib{R}'+\mib{r}_{\nu'}),
\label{eqn:bH=Ldm}
\end{equation}
where we use the indexing rule described in \S\ref{subsec:indexing} to express
 the site location.
The index $(\nu',\vR')$ runs
 over all nearest neighbors of the site $(\nu,\vR)$.
The elements of $4 \times 4$ matrix ${\it\Lambda}$ are given as
\begin{eqnarray}
&&
 {\it\Lambda}_{\nu\nu'}={\it\Lambda}_{\mbox{{\scriptsize off}}}+({\it\Lambda}_{\mbox{{\scriptsize diag}}}-{\it\Lambda}_{\mbox{{\scriptsize off}}})\delta_{\nu\nu'}\,,\label{eqn:Lambda}\\
&&
 {\it\Lambda}_{\mbox{{\scriptsize diag}}}
 = -\frac{\,1\,}{\,2\,}+\frac{\,Z d_1\,}{\,16\,}\,,\quad
 {\it\Lambda}_{\mbox{{\scriptsize off}}}
 = \frac{\,Z d_2\,}{\,16\,}
\,,
\end{eqnarray}
with
\begin{equation}
 d_1=\frac{1+3\eta-3\eta{}^2+3\eta{}^3}{2(1+\eta{}^3)}\,,\quad
 d_2=\frac{1-\eta+\eta{}^2-\eta{}^3}{2(1+\eta{}^3)}\,.
\end{equation}

In order to obtain the magnetic susceptibility from
 eq.(\ref{eqn:bH=Ldm}) we decompose it into the Fourier components by
  utilizing the translational vector $\vR$.
We define the Fourier
 transformation of ${\it\Delta}m$, $H$, ${\it\Lambda}_{\nu\nu'}$ for
  each sublattice as
\begin{eqnarray}
 &&\!\!\!\!\! {\it\Delta}m_{\mib{q};\nu}
 = {\rm e}^{-{\rm i} \mib{q}\cdot\mib{r}_{\nu}}
   \sum_{\mib{R}}
   \,{\rm e}^{-{\rm i} \mib{q}\cdot\mib{R}} {\it\Delta}m(\mib{R}+\mib{r}_{\nu})\,,\\
 &&\!\!\!\!\! H_{\mib{q};\nu}
 = {\rm e}^{-{\rm i} \mib{q}\cdot\mib{r}_{\nu}}
   \sum_{\mib{R}}
   \,{\rm e}^{-{\rm i} \mib{q}\cdot\mib{R}} H(\mib{R}+\mib{r}_{\nu})\,,\\
 &&\!\!\!\!\! {\it\Lambda}_{\mib{q};\nu\nu'}
 = 2 {\it\Lambda}_{\nu\nu'}\cos\left(\mib{q}\cdot(\mib{r}_{\nu}-\mib{r}_{\nu'})\right)\,.\label{eqn:Lambda_q}
\end{eqnarray}
Then eq.(\ref{eqn:bH=Ldm}) is reduced to the linear relation for
 sublattice magnetizations for each wave-number $\mib{q}$:
\begin{equation}
 {\it\Delta}m_{\mib{q};\nu}
 = \beta \sum_{\nu'} ({\it\Lambda}^{-1})_{\mib{q};\nu\nu'}\,H_{\mib{q};\nu'}\,
 = \sum_{\nu'} \chi_{\mib{q};\nu\nu'}\,H_{\mib{q};\nu'}\,,
\end{equation}
where $\chi_{\mib{q};\nu\nu'}=\beta ({\it\Lambda}^{-1})_{\mib{q};\nu\nu'}$
 is the wave-number dependent susceptibility.

\begin{figure}[tbp]
\begin{center}
\epsfile{file=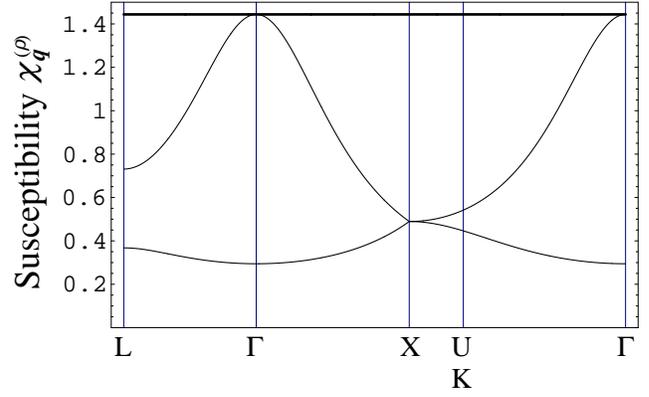,width=8.5cm}
\end{center}
\caption{
The $\mib{q}$-dependency of eigenvalues of $\chi_{\mib{q};\nu\nu'}$ in the
 first Brillouin zone along certain symmetry directions at $k_{\rm B}T/J=1$.
} 
\label{fig:q-4eigen_chi}
\end{figure}

Now all we need is to take the inverse of ${\it\Lambda}$-matrix with
 martix elements (\ref{eqn:Lambda_q}).
In the present case we can derive
 its eigenvalues analytically (see appendix 
\ref{appendix:eigenvalue}) and these are found to be 
\begin{eqnarray}
&&\lambda_{\mib{q}}^{(1)} = \lambda_{\mib{q}}^{(2)} = 2({\it\Lambda}_{\mbox{{\scriptsize diag}}}-{\it\Lambda}_{\mbox{{\scriptsize off}}})\,,\nonumber\\
&&\lambda_{\mib{q}}^{(3)} = 2({\it\Lambda}_{\mbox{{\scriptsize diag}}}+{\it\Lambda}_{\mbox{{\scriptsize off}}}) - {\it\Lambda}_{\mbox{{\scriptsize off}}}\sqrt{{\textstyle\sum_{\nu\nu'}}\cos\left(2 \mib{q}\cdot (\mib{r}_{\nu}-\mib{r}_{\nu'})\right)}\,.\nonumber\\
&&\lambda_{\mib{q}}^{(4)} = 2({\it\Lambda}_{\mbox{{\scriptsize diag}}}+{\it\Lambda}_{\mbox{{\scriptsize off}}}) + {\it\Lambda}_{\mbox{{\scriptsize off}}}\sqrt{{\textstyle\sum_{\nu\nu'}}\cos\left(2 \mib{q}\cdot (\mib{r}_{\nu}-\mib{r}_{\nu'})\right)}\,.\nonumber\\
\label{eqn:eigenvalue}
\end{eqnarray}
Note that
 the eigenvalues $\lambda_{\mib{q}}^{(1)}$ and $\lambda_{\mib{q}}^{(2)}$
  are degenerated and independent of $\mib{q}$.
Correspondingly the eigenvalues of $\chi_{\mib{q};\nu\nu'}$ are given as
\begin{equation}
 \chi_{\mib{q}}^{(\rho)}=\frac{\beta}{\lambda_{\mib{q}}^{(\rho)}},
\end{equation}
and $\chi_{\mib{q}}^{(1)}$ and $\chi_{\mib{q}}^{(2)}$ are the largest ones
 since an inequality
\begin{equation}
 0 < \lambda_{\mib{q}}^{(1)} = \lambda_{\mib{q}}^{(2)} \le \lambda_{\mib{q}}^{(3)} \le \lambda_{\mib{q}}^{(4)}
\label{eqn:4lambda}
\end{equation}
holds for any $\mib{q}$ and any $T$.
This proves the absence of magnetic phase transition
 in the present approximation
  because the largest eigenvalue of $\chi_{\mib{q};\nu\nu'}$
  is analytic and never diverges for any temperature $T$.
The $\mib{q}$-dependency of $\chi^{(\rho)}_{\mib{q}}$ are
 shown in Fig.~\ref{fig:q-4eigen_chi}.

Once all eigenvalues $\chi_{\mib{q}}^{(\rho)}$
 and the corresponding eigenvectors $u_{\mib{q};\nu}^{(\rho)}$ are known,
  the $\mib{q}$-dependent susceptibility matrix
   $\chi_{\mib{q};\nu\nu'}$ is expressed as
\begin{equation}
 \chi_{\mib{q};\nu\nu'} = \sum_\rho
   u_{\mib{q};\nu}^{(\rho)} \chi_{\mib{q}}^{(\rho)} u_{\mib{q};\nu'}^{(\rho)}.
\end{equation}
The final form of magnetization is thus obtained as follows:
\begin{equation}
  {\it\Delta}m_{\mib{q};\nu\alpha}
 = \sum_{\nu'\alpha'} \chi_{\mib{q};\nu\alpha;\nu'\alpha'}\,H_{\mib{q};\nu'\alpha'}\,.
\end{equation}
where
\begin{eqnarray}
 && {\it\Delta}m_{\mib{q};\nu\alpha}={\it\Delta}m_{\mib{q};\nu}n_{\nu\alpha}, \\
 && H_{\mib{q};\nu}={\textstyle\sum_\alpha} H_{\mib{q};\nu\alpha}n_{\nu\alpha}\,,\\
 && \chi_{\mib{q};\nu\alpha;\nu'\alpha'}=n_{\nu\alpha}\chi_{\mib{q};\nu\nu'}n_{\nu'\alpha'}\,,
\end{eqnarray}
and $n_{\nu\alpha}$ is the $\alpha\,(=x,\,y,\,z)$ component of $\mib{n}_\nu$.

\subsection{Neutron Scattering Intensity}
\begin{figure}[tbp]
\begin{center}
\epsfile{file=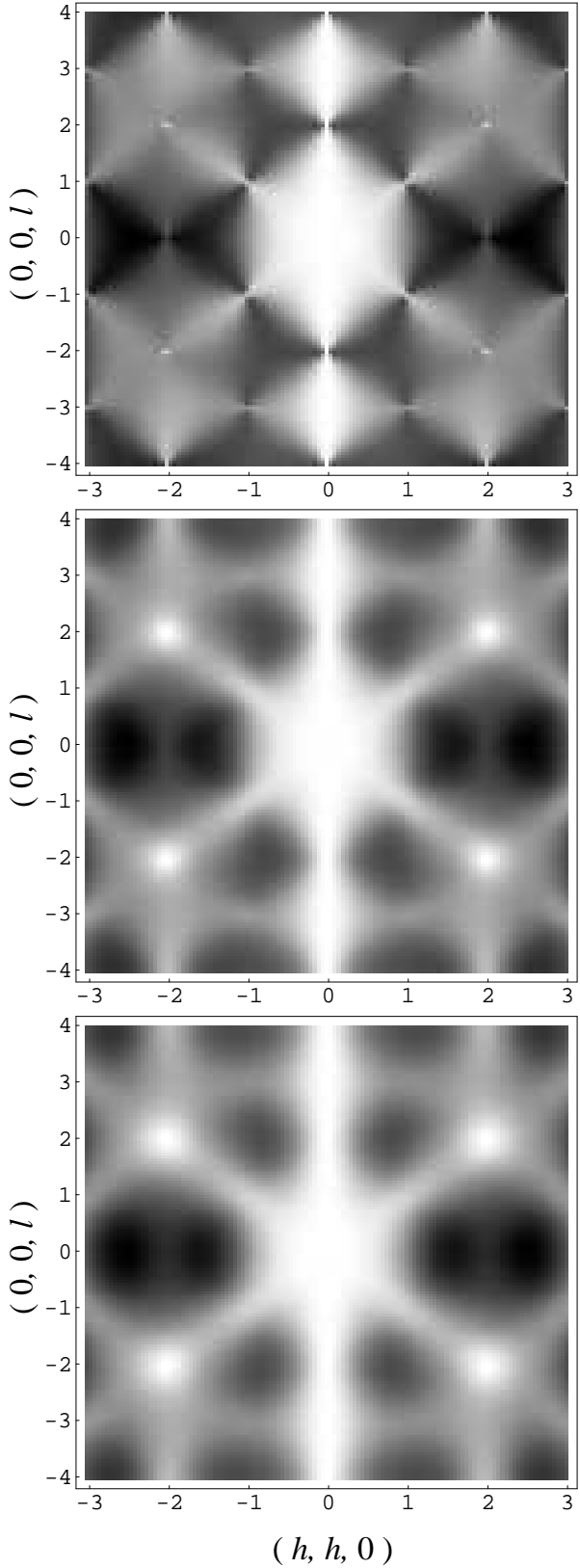,width=8cm}
\end{center}
\caption{
Neutron scattering pattern in the $(h\,h\,l)$ plane of reciprocal space
 at $k_{\rm B}T/J=0.1,\,0.5,\,1.0$ .
Black shows the lowest intensity level, white the heighest.
} 
\label{fig:S_hhl}
\end{figure}

As an application of the present result we evaluate the diffuse magnetic
 scattering intensity:
\begin{eqnarray}
\lefteqn{I(\,\mib{Q}=\mib{G}+\mib{q}\,)}\quad &&\nonumber\\
 &=& I_0 \left(f(\mib{Q})\right)^2 \sum_{\nu\alpha;\nu'\alpha'} \left(\delta_{\alpha\alpha'}-\frac{\,Q_{\alpha}Q_{\alpha'}\,}{\,Q{}^2\,}\right)\nonumber\\
&& \quad \times\chi_{\mib{q};\nu\alpha;\nu'\alpha'} \cos\left(\mib{G}\cdot(\mib{r}_{\nu}-\mib{r}_{\nu'})\right),
\end{eqnarray}
where $\mib{G}$, $I_0$ and $f(\mib{Q})$ are, respectively, a reciprocal
 vector, a constant and an atomic form factor. 
Numerical evaluation of the neutron scattering pattern with
 $f(\mib{Q})=1$ is shown in Fig.\ref{fig:S_hhl}. 
As temperature becomes higher, the calculated pattern becomes vaguer
 due to the fluctuation. 
Although almost the same scattering pattern is obtained from the MFA in
 the paramagnetic phase, the tetrahedral cactus approximation in the CVM
  shows vaguer pattern than the MFA does at the same temperature. 
This difference is due to the fluctuation caused by the four spin
 correlation which the tetrahedral cactus approximation in the CVM can
  take into account. 
The scattering pattern agrees qualitatively with the results of Monte
 Carlo simulation on the basis of the nearest neighbor spin ice
  model.~\cite{Bramwell2001} 

\section{Summary and Discussion}\label{sec:discussion}
We have applied the tetrahedral cactus approximation in the CVM to the
 spin ice system with nearest neighbor interactions.
The temperature dependence of the entropy and the specific heat
 shows qualitatively good agreement with those observed
  by Monte Carlo simulations and experiments, and the Pauling value
   is reproduced for the residual entropy.
We have obtained the analytic expression of
 the $\mib{q}$-dependent magnetic susceptibility and confirmed
  the absence of magnetic phase transition.
Furthermore we have evaluated the neutron scattering pattern,
 which is also consistent with that obtained from Monte Carlo simulations.

\begin{figure}[tbp]
\begin{center}
\epsfile{file=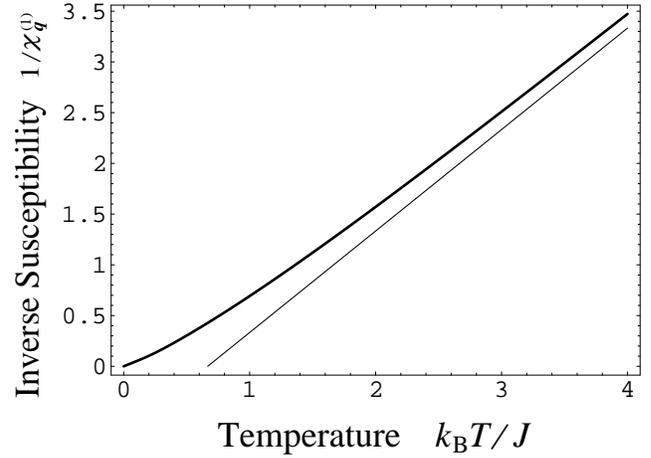,width=8.5cm}
\end{center}
\caption{
Inverse of the maximum eigenvalue $\chi_{\mib{q}}^{(1)}$ versus
 temperature. The thick and thin lines represent
 those obtained by the cactus approximation and the
 usual mean field approximation, respectively. 
}
\label{fig:inverse_chi}
\end{figure}

The crucial point which differentiates the present results from
 those of the MFA is that the tetrahedral cactus approximation can take
  into account the ice rule in an appropriate manner.
It is because that the four-spin correlation between the spins on a tetrahedron
 plays an important role for the ice rule of spin ice system
  in low temperatures.
To depict this situation let us explain briefly what results in the
 point approximation of the CVM, which corresponds to the MFA.
The approximant $\Ptot$ is given by eq.(\ref{eqn:point}) and 
 the  same procedure as in the previous
  sections leads to the ${\it\Lambda}$-matrix (\ref{eqn:Lambda})
   with the following parameters replaced by 
\begin{equation}
{\it\Lambda}_{\mbox{{\scriptsize diag}}} = \frac{\,1\,}{\,2\,}\,,\quad
{\it\Lambda}_{\mbox{{\scriptsize off}}} = \frac{\,\Jeff\,}{\,3k_{\rm B}T\,}\,.
\end{equation}
The rest of the story, including
eqs.(\ref{eqn:eigenvalue}) - (\ref{eqn:4lambda}), is also valid
 for the point approximation and we thus obtain $\chi_{\mib{q}}^{(\rho)}$.
The temperature dependence of the maximum eigenvalue $\chi_{\mib{q}}^{(1)}$ 
 is shown in Fig.\ref{fig:inverse_chi}. 
In this case it diverges for all $\mib{q}$ simultaneously
 at $k_{\rm B}T_{\rm c}/\Jeff = 2/3$ since $\chi_{\mib{q}}^{(1)}=\beta
 /(1-2\beta \Jeff/3)$ in the point approximation. 
Below $T_{\rm c}$ the long range order appears and prevents the system
 from having the expecting residual entropy.
Obviously this result is not suitable for 
 the behavior of the present model at low
  temperatures where the ice rule dominates.
 
Recently dipolar spin ice systems have been found
 to show a remarkable aspect.
It has been suggested that in the spin ice materials such as
 Dy$_2$Ti$_2$O$_7$ and Ho$_2$Ti$_2$O$_7$ the long range dipolar interaction
  is responsible for spin ice behaviors. 
It would be then expected 
 that the macroscopic degeneracy of the ground state is removed by the
  dipolar interaction and the long range order exists at very low
   temperature. 
The existence of ordered phase is also indicated by the Monte
 Carlo simulation by Melko {\it et al}.~\cite{Melko2000}
On the other hand,  such a phase has not been found in the real spin ice
 materials Dy$_2$Ti$_2$O$_7$ and Ho$_2$Ti$_2$O$_7$.~\cite{Bramwell2001} 
This may happen if with the temperature being decreased
 the free energy barriers separating the ordered state
  from quasi-degenerate states grow very high and consequently
   the relaxation time to the ordered phase becomes very large.
In order to elucidate the ordered phase of the dipolar spin
 ice more in detail, we have to cope with the long range dipolar
  interactions neglected in the present work.
The application of the cactus approximation to the dipolar spin ice is
 in progress and will be presented elsewhere in the near future.

\section{Acknowledgments}
The authors would like to thank Prof.~M.~Tokunaga for his encouragements
 and helpful comments, and Dr K.~Matsuhira for introducing us to spin
  ice systems.

\appendix
\section{Derivation of Eq.(\ref{eqn:bH=Ldm})}\label{appendix:bH=Ldm}  
First we show the extremum condition (\ref{eqn:dF/dm=0}) for the free
 energy more in detail.
For instance, the differentiation with respect to site magnetization $m_i$ 
 is written as 
\begin{eqnarray}
 \beta\frac{\partial F}{\partial m_i}
 &=& \frac{\,1\,}{\,2^4\,}\, \sum_{\langle jkl \rangle} \mbox{{\large\rm Tr}} \bigl[ \beta\,{\cal H}^{(ijkl)}_4(\sigma_i,\,\sigma_j,\,\sigma_k,\,\sigma_l)\, \sigma_i\bigr]\nonumber\\
 && - \frac{\,1\,}{\,2\,}\, \mbox{{\large\rm Tr}} \bigl[\sigma_i \ln P^{(i)}_1(\sigma_i)\bigr] \label{eqn:dF/dm}\\
 && + \frac{\,1\,}{\,2^4\,}\, \sum_{\langle jkl \rangle}  \mbox{{\large\rm Tr}} \bigl[\sigma_i \ln P^{(ijkl)}_4(\sigma_i,\,\sigma_j,\,\sigma_k,\,\sigma_l)\bigr]\,.\nonumber
\end{eqnarray}
The summation $\sum_{\langle jkl \rangle}$ is taken over two tetrahedra 
which includes the $i$-th site.
By keeping the response of order parameters up to the linear order of 
external field, eq.(\ref{eqn:dF/dm}) is written as
\begin{eqnarray}
\beta\frac{\partial F}{\partial m_i}
 &\simeq& -\beta H_i - {\it\Delta}m_i \label{eqn:dF/dm_i}\\[-2mm]
 && \quad + \sum_{\langle jkl \rangle}  \mbox{{\large\rm Tr}} \biggl[ \frac{\sigma_i {\it\Delta}P^{(ijkl)}_4(\sigma_i,\,\sigma_j,\,\sigma_k,\,\sigma_l) }{2^4\,P_4^0(\sigma_i,\,\sigma_j,\,\sigma_k,\,\sigma_l)} \biggr]\,.\nonumber
\end{eqnarray}
Similarly the rests of the eqs.(\ref{eqn:dF/dm=0}) are
 expanded up to the first order as
\begin{equation}
\beta\frac{\partial F}{\partial m_{ab}}
 \simeq \mbox{{\large\rm Tr}} \biggl[ \frac{\sigma_a\sigma_b {\it\Delta}P^{(ijkl)}_4(\sigma_i,\,\sigma_j,\,\sigma_k,\,\sigma_l) }{2^4\,P_4^0(\sigma_i,\,\sigma_j,\,\sigma_k,\,\sigma_l)} \biggr]=0\,, \\
\label{eqn:dF/dm2=0}
\end{equation}
\begin{equation}
\beta\frac{\partial F}{\partial m_{abc}}
 \simeq \mbox{{\large\rm Tr}} \biggl[ \frac{\sigma_a\sigma_b\sigma_c {\it\Delta}P^{(ijkl)}_4(\sigma_i,\,\sigma_j,\,\sigma_k,\,\sigma_l) }{2^4\,P_4^0(\sigma_i,\,\sigma_j,\,\sigma_k,\,\sigma_l)} \biggr]=0\,, \\
\label{eqn:dF/dm3=0}
\end{equation}
\begin{equation}
\beta\frac{\partial F}{\partial m_{ijkl}}
 \simeq \mbox{{\large\rm Tr}} \biggl[ \frac{\sigma_i\sigma_j\sigma_k\sigma_l {\it\Delta}P^{(ijkl)}_4(\sigma_i,\,\sigma_j,\,\sigma_k,\,\sigma_l) }{2^4\,P_4^0(\sigma_i,\,\sigma_j,\,\sigma_k,\,\sigma_l)} \biggr]=0\,. \\
\label{eqn:dF/dm4=0}
\end{equation}
where $\it\Delta$ signifies quantities of linear order to external field.
In the paramagnetic phase short range order parameters such as $m_{ij}$
 and $m_{ijkl}$ have no linear response because these are even functions
  of the field.
Then eqs.(\ref{eqn:dF/dm_i}) and (\ref{eqn:dF/dm3=0}) are expressed 
 as: 
\begin{equation}
\beta H_i = 
 - {\it\Delta}m_i + \frac{\,Z\,}{\,2^8\,} \sum_{\langle jkl \rangle} \left(\mib{M}_1^{\rm t}{\it\Delta}\mib{m}_1+\mib{M}_3^{\rm t}{\it\Delta}\mib{m}_3\right)\,,\label{eqn:bH=L1dm1+L3dm3}
\end{equation}
\begin{equation}
 W_3 {\it\Delta}\mib{m}_1 + W_1 {\it\Delta}\mib{m}_3 = 0\,,\label{eqn:M3dm1+M1dm3=0}
\end{equation}
where a superscript ``t'' denotes a transpose of a vector and
\begin{eqnarray}
&& \mib{m}_1 = (m_i,\,m_j,\,m_k,\,m_l)^{\rm t}\,,\quad\\
&& \mib{m}_3 = (m_{jkl},\,m_{kli},\,m_{lij},\,m_{ijk})^{\rm t}\,.
\end{eqnarray}
The 4-column vectors $\mib{M}_1$ and $\mib{M}_3$ are given by 
\begin{equation}
 \mib{M}_1^{\rm t}=(A,\,B,\,B,\,B)\,,\quad
 \mib{M}_3^{\rm t}=(C,\,B,\,B,\,B)\,
\end{equation}
and the 4$\times$4 matrices $W_1$ and
$W_3$ by
\begin{eqnarray}
&&
(W_1)_{ij}=A\delta_{ij}+B\,(1-\delta_{ij}),
\\
&&
(W_3)_{ij}=C\delta_{ij}+B\,(1-\delta_{ij})
\end{eqnarray}
with
\begin{eqnarray}
&& A
 = \mbox{\large\rm Tr}\left[\frac{1}{{\rm e}^{-\beta {\cal H}_4^0(\sigma_i,\,\sigma_j,\,\sigma_k,\,\sigma_l)}}\right]
 = 6 \eta + 8 + 2 \eta^{-3}\,,\nonumber\\ 
&& B
 = \mbox{\large\rm Tr}\left[\frac{\sigma_i \sigma_j}{{\rm e}^{-\beta {\cal H}_4^0(\sigma_i,\,\sigma_j,\,\sigma_k,\,\sigma_l)}}\right]
 = -2 \eta + 2 \eta^{-3}\,,\nonumber\\ 
&& C
 = \mbox{\large\rm Tr}\left[\frac{\sigma_i\sigma_j\sigma_k\sigma_l}{{\rm e}^{-\beta {\cal H}_4^0(\sigma_i,\,\sigma_j,\,\sigma_k,\,\sigma_l)}}\right]
 = 6 \eta - 8 + 2 \eta^{-3}\,.\nonumber\\ 
\end{eqnarray}
By eliminating ${\it\Delta}\mib{m}_3$ from
 eqs.(\ref{eqn:bH=L1dm1+L3dm3}) and (\ref{eqn:M3dm1+M1dm3=0}),
the final expression between magnetization and magnetic field is obtained as
\begin{eqnarray}
\beta H_i
 &=& 
 - {\it\Delta}m_i + \frac{\,Z\,}{\,2^8\,} \sum_{\langle jkl \rangle} \left(\mib{M}_1^{\rm t}+\mib{M}_3^{\rm t}(W_1)^{-1}W_3\right){\it\Delta}\mib{m}_1\nonumber\\
 &=& 
 \sum_{\langle jkl \rangle}\left({\it\Lambda}_{\mbox{\scriptsize diag}}{\it\Delta}m_i + {\it\Lambda}_{\mbox{\scriptsize off}}({\it\Delta}m_j+{\it\Delta}m_k+{\it\Delta}m_l)\right)\,.\nonumber\\[-3mm]
\end{eqnarray}
The above equation is easily rewritten
in terms of the indexing rule described in \S\ref{subsec:indexing}
to give eq.(\ref{eqn:bH=Ldm}).

\appendix
\section{Derivation of Eq.(\ref{eqn:eigenvalue})}\label{appendix:eigenvalue}
We express the ${\it\Lambda}$-matrix
(\ref{eqn:Lambda_q}) in the form as 
\begin{eqnarray}
{\it\Lambda}_{\mib{q};\nu\nu'}
 &=& 2 ({\it\Lambda}_{\mbox{{\scriptsize diag}}}
    - {\it\Lambda}_{\mbox{{\scriptsize off}}}) \delta_{\nu\nu'}
\nonumber\\
 && \qquad
 +2 {\it\Lambda}_{\mbox{{\scriptsize off}}}
  \cos\left(\mib{q}\cdot(\mib{r}_{\nu}-\mib{r}_{\nu'})\right)\,.
\end{eqnarray}
The problem is then reduced to obtaining the eigenvalues of the matrix 
$\cos\left(\mib{q}\cdot(\mib{r}_{\nu}-\mib{r}_{\nu'})\right)$: 
\begin{eqnarray}
\lefteqn{
 \cos\left(\mib{q}\cdot(\mib{r}_{\nu}-\mib{r}_{\nu'})\right)
}\quad &&\nonumber\\
&& = \cos(\mib{q}\cdot\mib{r}_{\nu})\cos(\mib{q}\cdot\mib{r}_{\nu'})
 + \sin(\mib{q}\cdot\mib{r}_{\nu})\sin(\mib{q}\cdot\mib{r}_{\nu'})
 \nonumber\\
&& = (\mib{c}\mib{c}^{\rm t}+\mib{s}\mib{s}^{\rm t})_{\nu\nu'}\,,
\end{eqnarray}
where
\begin{equation}
 \mib{c}=\left(\!\!\!
\begin{array}{c}
 \cos(\mib{q}\cdot\mib{r}_1)\\
 \cos(\mib{q}\cdot\mib{r}_2)\\
 \cos(\mib{q}\cdot\mib{r}_3)\\
 \cos(\mib{q}\cdot\mib{r}_4)\\
\end{array}
\!\!\!\right)\,,\quad
 \mib{s}=\left(\!\!\!
\begin{array}{c}
 \sin(\mib{q}\cdot\mib{r}_1)\\
 \sin(\mib{q}\cdot\mib{r}_2)\\
 \sin(\mib{q}\cdot\mib{r}_3)\\
 \sin(\mib{q}\cdot\mib{r}_4)\\
\end{array}
\!\!\!\right)\,.
\end{equation}
Thus the matrix under consideration is just a sum of projectors.
Then it is obvious that the subspace normal to $\mib{c}$ and
$\mib{s}$ is a two-dimensional eigenspace with the eigenvalue $0$. 
The rest of the eigenvectors with non-zero eigenvalue can be expressed
 by a linear combination of $\mib{c}$ and
$\mib{s}$ as $c_1\mib{c}+c_2\mib{s}$,
where $c_1$ and $c_2$ are chosen so as to satisfy 
\begin{equation}
 (\mib{c}\mib{c}^{\rm t}+\mib{s}\mib{s}^{\rm t})
(c_1 \mib{c}+c_2 \mib{s})
 = \kappa (c_1\mib{c}+c_2\mib{s})\,,
\end{equation}
where $\kappa$ is the eigenvalue. 
We can rewrite this equation as 
\begin{equation}
 \left(\!\!\!\begin{array}{cc}
 \mib{c}^{\rm t}\mib{c} & \mib{s}^{\rm t}\mib{c} \\
 \mib{s}^{\rm t}\mib{c} & \mib{s}^{\rm t}\mib{s} 
 \end{array}\!\!\!\right)
\left(\!\!\!\begin{array}{c}
 c_1\\
 c_2\\
\end{array}\!\!\!\right)
 = \kappa\left(\!\!\!\begin{array}{c}
 c_1\\
 c_2\\
\end{array}\!\!\!\right)\,.
\end{equation}
The secular equation is then written as
\begin{equation}
 \det\left(\!\!\!\begin{array}{cc}
 \mib{c}^{\rm t}\mib{c}-\kappa & \mib{s}^{\rm t}\mib{c} \\
 \mib{s}^{\rm t}\mib{c} & \mib{s}^{\rm t}\mib{s}-\kappa 
 \end{array}\!\!\!\right)=0\,,
\end{equation}
which can be solved easily to obtain
\begin{equation}
 \kappa = 2 \pm \frac{\,1\,}{\,2\,}\sqrt{{\textstyle\sum_{\nu\nu'}}\cos\left(2 \mib{q}\cdot (\mib{r}_{\nu}-\mib{r}_{\nu'})\right)}\,.
\end{equation}

\end{document}